\documentclass[a4paper,12pt]{article}
\pdfoutput=1

\usepackage{graphicx}
\usepackage{colordvi}
\usepackage{fancybox}
\usepackage{cancel}
\usepackage{amsmath}
\usepackage{amssymb}
\usepackage{wasysym}
\usepackage{hyperref}
\usepackage{enumerate}
\usepackage{gensymb}

\usepackage[compress,numbers,sort]{natbib}

\usepackage{color}
\definecolor{gray}{rgb}{0.5,0.5,0.5}

\def\eq#1{{Eq.~(\ref{#1})}}
\def\eqs#1#2{{Eqs.~(\ref{#1})--(\ref{#2})}}
\def\fig#1{{Fig.~\ref{#1}}}

\def\sect#1{{Sect.~\ref{#1}}}
\def\sects#1#2{{Sects.~\ref{#1}--\ref{#2}}}
\def\app#1{{Appendix~\ref{#1}}}

\def\Tr{\mbox{Tr}\,}

\newcommand\lsim{\mathrel{\rlap{\lower4pt\hbox{\hskip1pt$\sim$}}
    \raise1pt\hbox{$<$}}}
\newcommand\gsim{\mathrel{\rlap{\lower4pt\hbox{\hskip1pt$\sim$}}
    \raise1pt\hbox{$>$}}}

\newcommand{\beq}{\begin{equation}}
\newcommand{\eeq}{\end{equation}}
\newcommand{\bea}{\begin{eqnarray}}
\newcommand{\eea}{\end{eqnarray}}
\newcommand{\bem}{\begin{pmatrix}}
\newcommand{\eem}{\end{pmatrix}}

\begin{document}

\numberwithin{equation}{section}

\begin{flushright}
CETUP2013-033, NORDITA-2015-23
\end{flushright}

\bigskip

\begin{center}

{\LARGE\bf  R-parity violation in SU(5)}

\vspace{1cm}

\centerline{
\large
Borut Bajc$^{a,}$\footnote{borut.bajc@ijs.si} 
and
Luca Di Luzio$^{b,}$\footnote{luca.di.luzio@ge.infn.it}  
}
\vspace{0.5cm}
\centerline{$^{a}$ {\it J.\ Stefan Institute, 1000 Ljubljana, Slovenia}}
\centerline{$^{b}$ {\it Dipartimento di Fisica, Universit\`a di Genova and INFN, Sezione di Genova,}} 
\centerline{{\it via Dodecaneso 33, 16159 Genova, Italy}}

\end{center}

\bigskip

\begin{abstract}

We show that judiciously chosen R-parity violating terms in the minimal renormalizable 
supersymmetric SU(5) are able to correct all the phenomenologically wrong mass 
relations between down quarks and charged leptons. The model can accommodate 
neutrino masses as well. One of the most striking consequences is a large mixing 
between the electron and the Higgsino. We show that this can still be in accord with 
data in some regions of the parameter space and possibly falsified in future experiments. 

\end{abstract}

\clearpage

\tableofcontents

\section{Introduction and outline}

SU(5) is the minimal and the simplest among supersymmetric grand unified theories (GUTs) \cite{Dimopoulos:1981zb}. 
It is thus of particular interest to test it in detail. 
In this work we will stick to its minimal 
renormalizable version with three matter copies of $\overline{5} \oplus 10$ and Higgs 
supermultiplets in $5 \oplus \overline{5} \oplus 24$.  
In fact, by allowing either non-renormalizable operators or extra superfields, 
many new unknown parameters enter in the 
superpotential thus making the model unpredictive. 

Yet, it is well known that the minimal renormalizable SU(5) GUT 
suffers from two main drawbacks. 
First, it predicts the equality at the GUT scale of the down quark and charged 
lepton masses ($i=1,2,3$ runs over generations)
\beq
\label{mDzmE0}
\frac{m_D^i}{m_E^i}=1 \, ,
\eeq
where $(m_D^1,m_D^2,m_D^3) = (m_d,m_s,m_b)$ and $(m_E^1,m_E^2,m_E^3) = (m_e,m_\mu,m_\tau)$.  
While, running those masses from their $m_Z$ values up to the GUT scale and
assuming (as an example) low-scale minimal supersymmetric standard model (MSSM) and low $\tan{\beta}$, 
gives (see e.g.~\cite{Babu:2012pb})
\beq
\label{YDzYEGUT}
\left(\frac{m_d}{m_e}, \frac{m_s}{m_\mu}, \frac{m_b}{m_\tau}\right) 
\approx \left(2.6, 0.23, 0.81\right) \, . 
\eeq
The discrepancies are of 
order one, and so cannot be easily accounted for without changing the theory, for example 
its physical content. The second problem is the absence of neutrino masses, similarly as in 
the standard model (SM). 

The issue of charged fermion masses in minimal renormalizable SU(5) can be solved by large supersymmetry (susy) breaking threshold 
corrections \cite{Hall:1985dx,Hall:1993gn,Hempfling:1993kv,Pierce:1996zz,Borzumati:1999sp,Antusch:2008tf,Enkhbat:2009jt,Iskrzynski:2014zla,Anandakrishnan:2014qxa,Iskrzynski:2014sba}. The prize to pay, however, is large $A$-terms which make the MSSM vacuum 
metastable \cite{Casas:1995pd}. Also, this does not bring any new ingredient for the solution of the neutrino mass problem.

In this work, we want to take instead an orthogonal approach. We neglect the contribution of susy threshold corrections 
and investigate whether the fermion mass ratio problem can be fixed by 
R-parity violating (RPV) \cite{Barbier:2004ez} couplings in the SU(5) model. This idea has been first proposed 
long ago \cite{Smirnov:1995ey} (for some other works in this direction see for 
example \cite{Diaz:1998wz,Dreiner:1995hu}), but never systematically worked out. 
We will show that R-parity violation can correct all the wrong mass relations (\ref{mDzmE0}). 
This will immediately open up a solution also for the neutrino mass problem. 
Let us now briefly describe the idea, while the details will be worked out in the body of the paper.  

It has been long known that giving up the minimal field content and allowing for 
extra vector-like matter fields it is possible to correct the SU(5) fermion mass relations 
(for an incomplete list of references see \cite{Witten:1979nr,
Berezhiani:1985in,Davidson:1987mi,Hisano:1993uk,bb2,
Berezhiani:1995yk,Babu:1995hr,bb3,Barr:2003zx,Malinsky:2007qy,zurab,Oshimo:2009ia}, \cite{Babu:2012pb}).  
As we will show later, the mixing of $d^c_i$ ($L_i$) 
with an extra vector-like color triplet (weak doublet) by angle $\theta_D^i$ ($\theta_E^i$) changes relation (\ref{mDzmE0}) 
into
\beq
\label{mDzmE}
\frac{m_D^i}{m_E^i}=\frac{\cos{\theta_D^i}}{\cos{\theta_E^i}} \, .
\eeq
However, we do not want to enlarge the field content of the model. An obvious (and well known) candidate for 
a vector-like pair is provided in the MSSM by the two Higgs doublets with bilinear RPV 
terms \cite{Aulakh:1982yn,Hall:1983id,Ross:1984yg,Ellis:1984gi,Masiero:1990uj}. 
According to (\ref{mDzmE}) with $\theta_D^i=0$ the mass 
ratio can only increase, so bilinear R-parity violation can be useful in the MSSM just for the first generation (see \eq{YDzYEGUT}). 

The next logical possibility is to allow also for color triplets $d^c_i$ to mix with the heavy 
SU(5) partners of the MSSM Higgses. At first glance this idea looks hopeless, 
since the mixing would induce the trilinear RPV couplings $\lambda'$ and $\lambda''$ (cf.~the superpotential in \eq{WRPV})
from the SU(5) Yukawas after rotation. 
This would make the proton to decay too fast, since 
the $d=4$ proton decay amplitude is proportional to $\lambda'\lambda''$ and suppressed just by 
the susy breaking scale. 
Moreover, SU(5) symmetry at the renormalizable level predicts for the RPV trilinear couplings (before rotation)
\beq
\label{3lambda}
\lambda=\lambda'=\lambda'' \, ,
\eeq
so that it seems impossible to disentangle $\lambda'$ from $\lambda''$. 
However (\ref{3lambda}) is valid 
in the original (flavour) basis, and not in the mass eigenbasis. Since the rotation of 
quarks $d^c_i$ with the heavy color anti-triplet $\bar{T}$ makes the mass and flavour eigenbasis different, 
we can avoid (\ref{3lambda}). At this point, 
special care must be taken to cancel $\lambda''=0$, effectively preserving the baryon number 
below the GUT scale. This can be obtained by taking a very specific value of the trilinear RPV couplings. 
As we will see, the requirement of $\lambda''=0$ will uniquely determine the other trilinear RPV couplings as a function 
of the mixings.

From \eq{mDzmE} it is clear that only the relative misalignment between doublet and triplet rotations matters for the correction of the mass 
eigenvalues. Hence, we will take the additional simplifying assumption that at a given generation $i$ either the quark $\theta_D^i$ or the 
lepton $\theta_E^i$ angle contributes, but not both. This will then uniquely determine the mixings (i.e.~the angles $\theta_D^i$ and $\theta_E^i$). 
By comparing (\ref{YDzYEGUT}) with (\ref{mDzmE}) we conclude that $d^c$ quarks 
of the second and third generation will mix with the heavy triplet, while only the first generation lepton will 
require a mixing with the Higgs doublet. We will hence have 
\begin{equation}
\frac{m_d}{m_e} = \frac{1}{\cos{\theta_E^1}} \, , \qquad  
\frac{m_s}{m_\mu} = \cos{\theta_D^2} \, ,  \qquad
\frac{m_b}{m_\tau} = \cos{\theta_D^3} \, .
\end{equation}
In the conclusions we will shortly comment on what happens if we 
relax these assumptions.

The resulting model turns out to be very much constrained. Not only one needs to do more than the usual single 
doublet-triplet fine-tuning, the original choice of the trilinear couplings must also magically combine in order 
to project to vanishing baryon number violating couplings after triplet rotation. Also, 
large lepton number violating couplings will induce tree and loop order neutrino masses, which 
will typically be too large unless under special conditions. 
We will not even attempt to understand or explain all these fortuitous relations among model parameters. 
But we will (shamelessly) use such possibility whenever needed by experimental data. 
In order to accommodate all these constraints our soft terms will not be subject to 
SU(5) invariant boundary condition at the GUT scale. We will hence assume that susy breaking is mediated below the GUT 
scale (for more comments on that see \sect{conclusions}). 
This exercise must be thus interpreted as a purely phenomenological possibility in order to avoid various 
constraints already in the minimal SU(5) model, and not as a proposal for a theoretically attractive theory. 

In spite of this, or better, because of this, the model predicts a phenomenologically very interesting situation 
of a large mixing between the electron (neutrino) 
and the charged (neutral) Higgsino. The seemingly ad-hoc assumption of only quark or lepton 
mixing in the same generation will at this point help in avoiding strong phenomenological constraints due to large 
(order one) lepton number violating couplings present in the low-energy MSSM Lagrangian. 
In particular, we will see that the tiny 
neutrino masses predict in this scenario a fixed (negative) ratio between the wino and bino masses, provided they 
are not much larger than the sfermion masses. 
Finally, the same large RPV couplings only allow a slowly decaying gravitino lighter than about 10 MeV 
as a dark matter (DM) candidate.

The paper is organised as follows: in \sect{RPVSU5} we discuss the general structure of the RPV SU(5) model 
and show how RPV interactions can correct the wrong mass relations of the original SU(5) model.  
Most of \sect{pheno} is instead devoted to checking whether the required amount of R-parity violation 
is still allowed by data. In particular, we discuss proton decay bounds, 
electroweak symmetry breaking, neutrino masses, modifications of SM couplings to leptons, 
lepton number and lepton flavour violating processes and gravitino DM.  
We conclude in \sect{conclusions} by recalling the main predictions of the model, 
while more technical details on the diagonalization of the relevant mass matrices are collected in \app{pertdiag}.

\section{The RPV SU(5)}
\label{RPVSU5}

The field content of the minimal SU(5) model is given by $5$, $\overline{5}_\alpha$ ($\alpha=0,1,2,3$), $10_i$ ($i=1,2,3$) and $24$. 
The decomposition of the SU(5) supermultiplets under the SM gauge quantum numbers reads
\beq
5=
\bem
T \cr
H_u
\eem
\, , \quad
%\;\;\;,\;\;\;
\bar 5_\alpha=
\bem
\bar 3 \cr
\bar 2
\eem_\alpha
\, , \quad
%\;\;\;,\;\;\;
10_i=
\bem
\epsilon_3u^c & Q \cr
-Q^T & -\epsilon_2e^c
\eem_i \, , 
\eeq
where $\epsilon_3$ ($\epsilon_2$) schematically denotes the Levi-Civita tensor in the SU(3) (SU(2)) space
and for the adjoint (which also spontaneously breaks SU(5) into the SM gauge group)
\beq
\label{24VEV}
24=\left(V+\frac{\phi_{(1,1)_0}}{\sqrt{30}}\right)
\bem
2 & 0 \cr 
0 & -3
\eem+
\bem
\phi_{(8,1)_0} & \phi_{(\bar 3, 2)_{5/6}} \cr
\phi_{(3,\bar 2)_{-5/6}} & \phi_{(1,3)_0}  
\eem \, .
\eeq
The indices of $\phi$ stand for the SM gauge quantum numbers, while the part 
proportional to $V$ denotes the GUT vacuum expectation value (vev).
%(hence $\langle \phi\rangle =0$).
The most general renormalizable superpotential is
\begin{align}
\label{W}
W_{\rm{SU(5)}} &= \bar 5_\alpha\left(M_\alpha+\eta_\alpha 24 \right)5+
\frac{1}{2}\Lambda_{\alpha\beta k}\bar 5_\alpha\bar 5_\beta 10_k \nonumber \\
& + Y^{10}_{ij} 10_i 10_j 5
+ \frac{M_{24}}{2} \Tr 24^2 + \frac{\lambda_{24}}{3} \Tr 24^3 \, , 
\end{align}
where SU(5) contractions are understood. In particular, $\Lambda_{\alpha\beta k}=-\Lambda_{\beta\alpha k}$ and 
$Y^{10}_{ij} = Y^{10}_{ji}$. The usual R-parity conserving (RPC) case 
\begin{align}
\label{WRPC}
W^{\rm RPC}_{\rm{SU(5)}} &= \bar 5_0\left(M_0+\eta_0 24 \right)5+
Y^5_{jk} \bar 5_0 \bar 5_j 10_k \nonumber \\
& + Y^{10}_{ij} 10_i 10_j 5
+ \frac{M_{24}}{2} \Tr 24^2 + \frac{\lambda_{24}}{3} \Tr 24^3 \, , 
\end{align}
is recovered in the limit 
\begin{equation}
M_\alpha = M_0 \delta_{\alpha0} \, , \qquad 
\eta_\alpha = \eta_0 \delta_{\alpha0} \, , \qquad
\Lambda_{\alpha\beta k} = Y^5_{jk} (\delta_{\alpha 0} \delta_{\beta j} -  \delta_{\alpha j} \delta_{\beta 0} ) \, .
\end{equation}

The terms in the second line of \eq{WRPC} and \eq{W} coincide: $Y^{10}$ is responsible for the up-quark masses, while $M_{24}$ and $\lambda_{24}$ 
participate to the GUT symmetry breaking and are related by the minimum equation to the SU(5) breaking vev in \eq{24VEV} 
via the relation $V = M_{24} / \lambda_{24}$. 
Moreover, in the RPC case $Y^5$ leads to the usual Yukawa unification condition (\ref{mDzmE0}) 
which we want to correct with the more general superpotential in (\ref{W}).  

Let us now focus on the first line of \eq{W}. 
From the first term we see that one combination of $\bar 5_\alpha$ gets a vector-like mass with $5$. 
Physically we know that such a mass has to be large in the triplet sector and 
light in the doublet one. This can be achieved thanks to SU(5) breaking via the vev contribution in \eq{24VEV}. 
Then the mass terms in the doublet-triplet sector 
of the superpotential become
\beq
\label{Wmass}
W_{\rm{mass}}=\bar 3_\alpha{\cal M}_\alpha T+\bar 2_\alpha\mu_\alpha H_u \, ,
\eeq
where
\bea
\label{triplet}
{\cal M}_\alpha&=&M_\alpha+2\eta_\alpha V \, , \\
\label{doublet}
\mu_\alpha&=&M_\alpha-3\eta_\alpha V \, .
\eea
The doublet-triplet splitting (assuming low-energy susy) means the following: 
\beq
\label{mualpha}
\mu_\alpha \lesssim {\cal O}(m_W) \, ,
\eeq
for all $\alpha=0,1,2,3$, while 
\beq
\label{Malpha}
{\cal M}_\alpha={\cal O}(M_{GUT}) \, ,
\eeq
for at least one $\alpha$. 
Notice that while in the usual RPC case one fine-tuning is enough, 
in the generic RPV case four fine-tunings are needed in order to satisfy \eq{mualpha} 
for all four possible choices of $\alpha$. 

Finally, the terms in $\Lambda_{\alpha\beta k}$ contain, on top of the above mentioned Yukawas, the 
trilinear RPV couplings which will be discussed in \sect{trilinearRPV}.

\subsection{The issue of the doublet basis}

Since in this setup there is no real difference between the four doublet superfields 
$\bar 2_\alpha=(N_\alpha,E_\alpha)^T$, what do we mean by the names 
(s)neutrino, charged or neutral Higgs(ino) and charged (s)lepton? In other words, what is the 
difference between neutral Higgs--sneutrino, neutral Higgsino--neutrino, 
charged Higgs--slepton and charged Higgsino--charged lepton? 
Although the results can always be written in a basis-independent way 
\cite{Davidson:2000uc,Davidson:2000ne} and so these names are strictly speaking 
not really necessary, we will still define such names for the sake of clarity. 

We will choose a convenient basis, in which only one among the SM doublets 
$\bar 2_\alpha\subset\bar 5_\alpha$ (let it be the one 
with index $\alpha = 0$) gets a nonzero vev $v_d$. This can be obtained by 
an SU(4) rotation of the $\bar 5_\alpha$ which affects the relations (\ref{triplet})-(\ref{doublet}) as well. 
One could argue that the new, rotated, $M_\alpha$ and $\eta_\alpha$ cannot be completely 
arbitrary, since the vevs themselves depend on them. However, it is not hard to imagine 
(and we will show it in more detail in \sect{EWSB}) that the freedom in the choice of soft terms 
allows us to consider $M_\alpha$ and $\eta_\alpha$ arbitrary with $\langle \bar 5_i\rangle=0$. 
Since we will not employ any particular spectrum of the soft terms, this is what we can 
(and will) do.

In particular, there are essentially four classes of fields we have to specify: the neutral 
bosons, the neutral fermions, the charged bosons and the charged fermions. 
These are fixed in the following way:

\begin{itemize}

\item
The flavour basis of {\bf neutral bosons} is defined such that 
the sneutrinos' vevs vanish:
\beq
\label{nuvev}
\langle\tilde\nu_i\rangle=0
%\;\;\;,\;\;\;
\, , \quad
i=1,2,3
\eeq
i.e.~we define the neutral Higgs vevs as in the RPC case:
\beq
\langle H_u^0\rangle\equiv v_u=v\sin{\beta}
\, , \quad
%\;\;\;,\;\;\;
\label{vd}
\langle H_d^0\rangle\equiv v_d=v\cos{\beta} \, , 
\eeq
where $v = 246$ GeV. More details about the electroweak symmetry breaking sector and the composition of the lightest Higgs boson 
in terms of the flavour basis can be found in \sect{EWSB}. 

\item
The {\bf neutral fermion} mass matrix is incorporated into the neutralino 
quadratic part of the lagrangian (see e.g.~\cite{Bisset:1998bt}):
\begin{multline}
\label{LN}
{\cal L}_N=-\frac{1}{2}
\bem
\tilde B^0 & \tilde W^0 & \tilde H_u^0 & \tilde H_d^0 & \nu_i
\eem
\\
\bem
M_1 & 0 & g'v_u/2 & -g'v_d/2 & 0 \cr
0 & M_2 & -gv_u/2 & gv_d/2 & 0 \cr
g'v_u/2 & -g v_u/2 & 0 & - \mu_0 & - \mu_k \cr
-g'v_d/2 & g v_d/2 & - \mu_0 & \eta_0\eta_0v_u^2/M_{\text{seesaw}} &\eta_0\eta_kv_u^2/M_{\text{seesaw}} \cr
0 & 0 & - \mu_i & \eta_i\eta_0v_u^2/M_{\text{seesaw}} & \eta_i\eta_kv_u^2/M_{\text{seesaw}}
\eem
\bem
\tilde B^0 \cr
\tilde W^0 \cr
\tilde H_u^0 \cr
\tilde H_d^0 \cr
\nu_k
\eem \, , 
\end{multline}
where we added the $4 \times 4$ lower-right block as the seesaw contribution from the SM 
singlet $(1,1)_0$ \cite{Minkowski:1977sc,Yanagida,gellmannramondslansky,Glashow,Mohapatra:1979ia} 
and weak triplet $(1,3)_0$ \cite{Foot:1988aq} states living in $24$, and
\beq
\label{1overM}
\frac{1}{M_{\text{seesaw}}}=\frac{3}{10}\frac{1}{M_{(1,1)_0}} + \frac{1}{2}\frac{1}{M_{(1,3)_0}} = -\frac{2}{5}\frac{1}{M_{24}} \, , 
\eeq
with $M_{24}$ denoting the superpotential parameter defined in \eq{W}, 
$M_{(1,3)_0}=-5M_{24}$ and $M_{(1,1)_0}=-M_{24}$.

It is clear from (\ref{LN}) that in the flavour basis $\tilde H_d^0$ is 
the fermionic superpartner of $H_d^0$ that gets the vev in (\ref{vd}). 
The mass basis is obtained by diagonalizing the matrix in \eq{LN} and 
neutrinos are the three lightest eigenstates.

\item
The {\bf charged fermions} are part of the chargino sector (see e.g.~\cite{Bisset:1998bt}):
\beq
\label{C}
{\cal L}_C=-
\bem
\tilde W^- & \tilde H_d^- & e_i
\eem
\bem
M_2 & gv_u/\sqrt{2} & 0 \cr
g v_d/\sqrt{2} & \mu_0 & 0 \cr
0 & \mu_i & \Lambda_{0ik}v_d
\eem
\bem
\tilde W^+ \cr
\tilde H_u^+ \cr
e_k^c
\eem \, .
\eeq
$\tilde H_d^-$ and $e_i$ are the weak partners of the previously defined $\tilde H_d^0$ and 
$\nu_i$, respectively.
In particular, the charged lepton mass eigenstates correspond to the three lightest eigenvalues of the matrix in \eq{C}.

\item
Finally the {\bf charged bosons}: in the flavour basis they are just the SU(2) partners of 
the neutral bosons defined through (\ref{nuvev}) and (\ref{vd}), or, equivalently, the bosonic 
superpartners of the charged fermions defined in (\ref{C}). We will denote them by $H_d^-$ 
and $\tilde e_i$.

\end{itemize}

This quadratic part of the Lagrangian, plus the analogous one for color triplets in (\ref{L3}), 
is RPC if ${\cal M}_i=\mu_i=0$. Of course, 
the whole Lagrangian, or even this part of it at higher loops, is not RPC due to nonzero trilinear 
terms, but in the basis we use, $\langle\tilde \nu_i\rangle=0$, these trilinear terms do not appear 
in the mass matrices at the tree order.

At this point, we are still free to rotate in the $3\times3$ subspace and we use this freedom to diagonalize 
the sub-matrix matrix
\beq
\label{baza}
\Lambda_{0ik}=\delta_{ik}d_k \, .
\eeq
Consequently, \eqs{triplet}{doublet} get rotated as well, but we will not keep track of it.

\subsection{The color triplet mass eigenstates}

The mass matrix for {\bf color triplets} comes from the first term in (\ref{Wmass}) and the last term in the first line of (\ref{W})
\beq
\label{L3}
{\cal L}_3=-
\bem
\bar 3_0 & \bar 3_i
\eem
\bem
{\cal M}_0 & 0 \cr
{\cal M}_i & \Lambda_{0ik}v_d
\eem
\bem
T \cr
Q_k
\eem \, . 
\eeq
The states $\bar 3_\alpha$ are still in the flavour basis. Let us rotate them 
into the mass eigenstates $(\bar T,d^c_k)$. 
Since ${\cal M}_\alpha={\cal O}(M_{GUT})\gg\Lambda_{0ik}v_d={\cal O}(m_W)$, we can easily disentangle 
the single heavy state $\bar T$ from the light ones $d^c_k$:
\beq
\bar 3_\alpha=
\bem
\bar T & d^c
\eem_\beta
U_{\beta\alpha} \, ,
\eeq
where the matrix $U$ projects the triplet states into the heavy direction
\bea
U(1,x_i)^T&\propto&(1,0,0,0)^T \, , \\
x_i&=&{\cal M}_i/{\cal M}_0 \, .
\eea
Assuming everything real for simplicity we have (see for example \cite{Barr:2003zx})
%\beq
%\label{U}
%U=
%\bem
%\bar\Lambda & x^T\Lambda \cr
%-\Lambda x & \Lambda
%\eem \, ,
%\eeq
\beq
\label{U}
U=
\bem
U_{00} & U_{0j} \cr
U_{i0} & U_{ij}
\eem \, ,
\eeq
where
%\bea
%\Lambda&=&1-\frac{xx^T}{\sqrt{1+x^Tx}\left(\sqrt{1+x^Tx}+1\right)} \, , \\
%\bar\Lambda&=&\frac{1}{\sqrt{1+x^Tx}} \, .
%\eea
%Or, in components:
\bea
\label{U00}
U_{00} &=& \frac{1}{\sqrt{1+\vec{x}^2}} \, , \\ 
\label{U0j}
U_{0j}&=&\frac{x_j}{\sqrt{1+\vec{x}^2}} \, , \\
\label{Ui0}
U_{i0}&=&-\frac{x_i}{\sqrt{1+\vec{x}^2}} \, , \\
\label{Uij}
U_{ij}&=&\delta_{ij}-\frac{x_ix_j}{\sqrt{1+\vec{x}^2}\left(\sqrt{1+\vec{x}^2}+1\right)} \, .
\eea
Then the light $3\times 3$ mass matrix (of the down quarks) is 
\beq
\label{MDx}
\left( M_D\right)_{ij}=U_{ij}(x)d_jv_d \, .
\eeq
Notice that $U_{ij}$ is \emph{not} unitary, since it is just the $3 \times 3$ sub-matrix of the $4 \times 4$ unitary $U_{\alpha\beta}$. 
This implies that the mass eigenvalue 
\begin{equation}
m_D^j (x) \leq m_D^j (0) = d_jv_d \, ,   
\end{equation}
for any $\vec{x}$.

\subsection{The charged lepton mass eigenstates}
\label{weakdec}

In order to get the three lightest eigenvalues of the chargino mass matrix 
it turns out to be a good approximation to consider the gaugino decoupling limit. 
This will be numerically confirmed in \sect{numex}. 
In this case what remains in \eq{C} is 
\beq
\bem
\mu_0 & 0 \cr
\mu_i & \Lambda_{0ik}v_d
\eem \, , 
\eeq
which is analogous to (\ref{L3}). Although the Higgsino mass is presumably much lighter than the 
GUT scale, it is still much heavier than the light charged leptons, so a similar rotation as in the case of the triplets 
can be used to integrate out the heavy Higgsino. The light charged lepton mass matrix is thus in this limit
\beq
\label{MEy}
\left(M_E\right)_{ij}=U_{ij}(y)d_jv_d \, , 
\eeq
with
\beq
y_i=\mu_i/\mu_0 \, .
\eeq
As before, we have 
\begin{equation}
m_E^j (x) \leq m_E^j (0) = d_jv_d \, ,   
\end{equation}
for any $\vec{y}$.

\subsection{How to avoid Yukawa unification}

We are interested in the correlation between down quarks (\eq{MDx}) and 
charged leptons (\eq{MEy}). 
It is known, see for example \cite{Babu:2012pb} and references therein, that with arbitrary $x_i$, $y_i$ and $d_i$, one can 
fit all down quark and charged lepton masses. In fact, defining the Yukawa $\lambda=m/v_d$, 
one finds in the hierarchical limit $d_1\ll d_2\ll d_3$ 
\bea
\label{yde}
\lambda_d=\frac{1}{\sqrt{1+x_1^2}}d_1\, , &&
\lambda_e=\frac{1}{\sqrt{1+y_1^2}}d_1\, , \\
\label{ysmu}
\lambda_s=\frac{\sqrt{1+x_1^2}}{\sqrt{1+x_1^2+x_2^2}}d_2\, , &&
\lambda_\mu=\frac{\sqrt{1+y_1^2}}{\sqrt{1+y_1^2+y_2^2}}d_2\, , \\
\label{ybtau}
\lambda_b=\frac{\sqrt{1+x_1^2+x_2^2}}{\sqrt{1+x_1^2+x_2^2+x_3^2}}d_3 \, , &&
\lambda_\tau=\frac{\sqrt{1+y_1^2+y_2^2}}{\sqrt{1+y_1^2+y_2^2+y_3^2}}d_3 \, .
\eea
From these equations it is clear that the most economical way to get (\ref{YDzYEGUT}) 
is to take $x_1=0$ (no mixing of the heavy color triplet 
with the first generation down quark) and 
$y_2=y_3=0$ (no mixing of the Higgsino with the second and third generation lepton). 

Before ending, we want to make a connection with the notation of \eq{mDzmE}. This can be done by defining the angles
\bea
\tan{\theta^1_D}=x_1\, , &&
\tan{\theta^1_E}=y_1\, , \\
\tan{\theta^2_D}=\frac{x_2}{\sqrt{1+x_1^2}}\, , &&
\tan{\theta^2_E}=\frac{y_2}{\sqrt{1+y_1^2}}\, ,\\
\tan{\theta^3_D}=\frac{x_3}{\sqrt{1+x_1^2+x_2^2}}\, , &&
\tan{\theta^3_E}=\frac{y_3}{\sqrt{1+y_1^2+y_2^2}}\, .
\eea
Then the masses are 
\beq
m^i_{D,E}=v_dd_i\cos{\theta^i_{D,E}}\, ,
\eeq
from which \eq{mDzmE} follows. 
%\beq
%\frac{m^i_D}{m^i_E}=\frac{\cos{\theta^i_D}}{\cos{\theta^i_E}}\, .
%\eeq

\subsection{A numerical example}
\label{numex}

As a numerical benchmark let us consider the case of MSSM with $\tan{\beta}=7$ and 
low susy scale. From the experimental values at $m_Z$ one can use the renormalization group equations (RGEs) 
to get the charged lepton and down quark Yukawa couplings at the GUT scale \cite{Babu:2012pb}
\bea
\label{YEGUT}
(\lambda^{\rm{exp}}_e, \lambda^{\rm{exp}}_\mu, \lambda^{\rm{exp}}_\tau) &=& (0.000013, 0.0028, 0.047) \, , \\
\label{YDGUT}
(\lambda^{\rm{exp}}_d, \lambda^{\rm{exp}}_s, \lambda^{\rm{exp}}_b) &=& (0.000034, 0.00063, 0.038) \, .
\eea

As we saw in the previous paragraph, the Yukawas can only diminish if a mixing 
with an extra vector-like $L-\overline{L}$ or $d^c-\overline{d^c}$ is introduced. Since 
from (\ref{YEGUT})-(\ref{YDGUT}) $\lambda_e^{\rm{exp}}<\lambda_d^{\rm{exp}}$, but $\lambda_\mu^{\rm{exp}}>\lambda_s^{\rm{exp}}$ 
and $\lambda_\tau^{\rm{exp}}>\lambda_b^{\rm{exp}}$, 
the minimal option is to keep $\lambda_d$, $\lambda_\mu$ and $\lambda_\tau$ unaltered, i.e.
\bea
d_1&=&\lambda_d=\lambda_d^{\rm{exp}}=0.000034 \, ,\\
d_2&=&\lambda_\mu=\lambda_\mu^{\rm{exp}}=0.0028 \, ,\\
d_3&=&\lambda_\tau=\lambda_\tau^{\rm{exp}}=0.047 \, ,
\eea
but correct (diminish) $\lambda_e=d_1$, $\lambda_s=d_2$, $\lambda_b=d_3$ to 
$\lambda_e^{\rm{exp}}$, $\lambda_s^{\rm{exp}}$, $\lambda_b^{\rm{exp}}$, respectively, by properly choosing 
the various $x_i$, $y_i$ (see \eqs{yde}{ybtau}):
\bea
\label{x1}
x_1&=&0 \, , \\
\label{x2}
x_2&=&\sqrt{(\lambda_\mu^{\rm{exp}}/\lambda_s^{\rm{exp}})^2-1}=4.3 \, , \\
\label{x3}
x_3&=&(\lambda_\mu^{\rm{exp}}/\lambda_s^{\rm{exp}})\sqrt{(\lambda_\tau^{\rm{exp}}/\lambda_b^{\rm{exp}})^2-1}=3.2 \, , \\
\label{y1}
y_1&=&\sqrt{(\lambda_d^{\rm{exp}}/\lambda_e^{\rm{exp}})^2-1}=2.4 \, , \\
\label{y23}
y_{2,3}&=&0  \, .
\eea

Notice that we fit all the masses at $M_{GUT}$. Although this is a correct procedure for 
the quarks, since we are integrating out the heavy (GUT scale) color triplet, the lepton (electron) corrections 
should be determined in principle at low energy, when the Higgsino is integrated out. 
But since the RGEs for the light Yukawas are essentially linear ($d\lambda_e/dt\propto \lambda_e$), the 
result is practically the same.

As a final remark, the r.h.s.~of \eq{yde} for the electron mass is only approximate, since 
the full mass matrix in \eq{C} contains mixings with gauginos as well. It is easy to check its consistency. 
The result is that the error by taking the approximate 
formula (\ref{yde}) is always below $2\%$ for $M_2 > 1$ TeV. 

\subsection{The trilinear RPV couplings}
\label{trilinearRPV}

Let us define the RPV superpotential of the low-energy MSSM effective theory as
\beq
\label{WRPV}
W_{\rm{RPV}}=H_u\mu_iL_i+\frac{1}{2}\lambda''_{ijk}d_i^cd_j^cu_k^c+\frac{1}{2}\lambda_{ijk}L_iL_je_k^c
+\lambda'_{ijk}d_i^cL_jQ_k \, . 
\eeq
The trilinear RPV couplings are then obtained by decomposing the SU(5) superpotential (\ref{W}) 
under the SM group and by matching it with \eq{WRPV}. This operation yields
\bea
\lambda''_{ijk}&=&U_{i\alpha}U_{j\beta}\Lambda_{\alpha\beta k} \, , \\
\lambda'_{ijk}&=&U_{i\alpha}\Lambda_{\alpha j k} \, , \\
\lambda_{ijk}&=&\Lambda_{ijk} \, .
\eea
By enforcing the safe condition\footnote{The exact condition $\lambda''=0$ is, strictly speaking, unnecessary. 
However, the most conservative bounds from matter stability require 
$|\lambda' \lambda''| < 10^{-10}$ for any flavour index 
and for superpartners around the TeV scale \cite{Smirnov:1996bg}, while analogous bounds 
hold as well for the combinations $|\lambda \lambda''|$ and $|\mu_i/\mu_0 \lambda''|$. 
Hence, in practice, large mixings in the triplet ($x_i$) and doublet ($y_i$) sectors require 
$\lambda'' \approx 0$. For a recent discussion of baryonic R-parity violation in GUTs see e.g.~\cite{DiLuzio:2013ysa}.}
\beq
\label{lambdaseq0}
0=\lambda''_{ijk}=(U_{i0}U_{jn}-U_{in}U_{j0})\Lambda_{0nk}+U_{il}U_{jn}\Lambda_{lnk}  \, ,
\eeq
we can calculate the other trilinear couplings. 
To this end, we use the choice of basis in (\ref{baza}), the explicit form of $U$ in (\ref{U}) and the relation
\beq
\left(\delta_{ik}+\frac{x_ix_k}{1+\sqrt{1+\vec{x}^2}}\right)
\left(\delta_{kj}-\frac{x_kx_j}{\sqrt{1+\vec{x}^2}(1+\sqrt{1+\vec{x}^2})}\right)=\delta_{ij}  \, ,
\eeq
which allows to compute the inverse of $U$. 
Hence, after some algebra we obtain
\beq
\lambda_{ijk}=(x_i\delta_{jk}-x_j\delta_{ik})d_k \, ,
\eeq
or explicitly (for the numerical example discussed in \sect{numex})
\bea
\label{lam3num}
\lambda_{ij3}&=&d_3
\bem
0 & 0 & x_1 \cr
0 & 0 & x_2 \cr
-x_1 & -x_2 & 0
\eem_{ij} \to 
\bem
0 & 0 & 0 \cr
0 & 0 & 0.20 \cr
0 & -0.20 & 0
\eem_{ij} \, ,
\\
\label{lam2num}
\lambda_{ij2}&=&d_2
\bem
0 & x_1 & 0 \cr
-x_1 & 0 & -x_3 \cr
0 & x_3 & 0
\eem_{ij}\to
\bem
0 & 0 & 0 \cr
0 & 0 & -0.0088 \cr
0 & 0.0088 & 0
\eem_{ij} \, ,
\\
\label{lam1num}
\lambda_{ij1}&=&d_1
\bem
0 & -x_2 & -x_3 \cr
x_2 & 0 & 0 \cr
x_3 & 0 & 0
\eem_{ij} \to
\bem
0 & -0.00014 & -0.00011 \cr
0.00014 & 0 & 0 \cr
0.00011 & 0 & 0
\eem_{ij} \, ,
\eea
where we used for our fit $x_1=0$. The only relevant matrix 
element (i.e.~$\propto d_3 = \lambda_\tau$) is then $\lambda_{233}=-\lambda_{323}$.

Similarly, for the other trilinear term we get 
\beq
\lambda'_{ijk}=\left(-x_j\delta_{ik}+
\frac{x_ix_jx_k}{\sqrt{1+\vec{x}^2}(1+\sqrt{1+\vec{x}^2})}\right)d_k \, .
\eeq
Even in this case the piece proportional to $\lambda_\tau$ never goes
through the first generation, i.e.~$\lambda'_{ijk}\propto \lambda_e$ if any among $i,j,k$ equals 1, since $x_1=0$.
This is important, since in this way many dangerous processes, like for example 
neutrinoless double $\beta$ decay, get automatically suppressed (cf.~\sects{LNvp}{LFVsect}). 
Numerically we get
\bea
\label{lamp3num}
\lambda'_{i3k}&\to&
\bem
-0.00011 &0 & 0 \cr
0 & -0.0042 & 0.059 \cr
0 & 0.0035 & -0.11
\eem_{ik} \, ,
\\
\label{lamp2num}
\lambda'_{i2k}&\to&
\bem
-0.00014 & 0 & 0 \cr
0 & -0.0056 & 0.079 \cr
0 & 0.0046 & -0.14
\eem_{ik} \, ,
\\
\label{lamp1num}
\lambda'_{i1k}&\to&
\bem
0 & 0 & 0 \cr
0 & 0 & 0 \cr
0 & 0 & 0
\eem_{ik} \, .
\eea

To summarize, the $L_1$ lepton number is strongly broken by the ${\cal O}(1)$ parameter $\mu_1/\mu_0$, 
$L_2$ by the ${\cal O}(0.1)$ couplings $\lambda_{233}$ and $\lambda'_{i23}$, $i=2,3$, 
and $L_3$ by the ${\cal O}(0.1)$ values of $\lambda'_{i33}$, $i=2,3$. 
Neutrino masses are thus generically expected to be large (see \sect{numasses}).
On the other hand, baryon number is effectively preserved below the GUT scale, 
thanks to the condition $\lambda''=0$.

\section{Phenomenology}
\label{pheno}

To study the phenomenology, we have to define our low-energy effective theory 
which is the MSSM with specific RPV couplings. 
As we saw, the low-energy RPV parameters considered so far are strongly correlated. In general they are 
parametrized by $x_i(={\cal M}_i/{\cal M}_0)$ and $y_i (=\mu_i/\mu_0)$. In order to simplify our analysis and minimize 
the corrections to be done, we assumed that the RPV parameters which make the fermion mass problem more severe are 
not present (i.e.~$x_{1}=y_{2,3}=0$). 
Due to that we will limit our phenomenological analysis to the case  
\begin{equation}
\vec{\mu} = (\mu_1, 0, 0) \, .
\end{equation}
Recall that the numerical values of the RPV couplings $\mu_i/\mu_0$, $\lambda_{ijk}$ and $\lambda'_{ijk}$ 
are at this point all known (cf.~\eqs{x1}{y23}, \eqs{lam3num}{lam1num} and \eqs{lamp3num}{lamp1num}), being determined by the requirement of vanishing baryon number violating couplings in the MSSM 
and the correct fit to fermion masses. 

To study the phenomenological consequences of the model we also need to specify the other 
RPV couplings which did not enter in the analysis so far, but which can still have a strong phenomenological 
impact: the soft mass terms $B_i$, $m_{0j}^2$ as well as the trilinears $A_{ijk}$, $A'_{ijk}$ and $A''_{ijk}$. 
Since it is not our intent to do here a full phenomenological study of the most general case, but 
just to show the existence of a realistic model, we will take further simplifying assumptions: let 

\begin{itemize}

\item
the RPV bilinear soft terms point in the direction 1, similarly as the $\mu_i$ in the superpotential
\bea
\label{Bi}
B_i&\propto&\delta_{i1} \, , \\
\label{m0i}
m_{0i}^2&\propto&\delta_{i1} \, .
\eea
Although one would be tempted to make both 
r.h.s.~in (\ref{Bi}) and (\ref{m0i}) to vanish, electroweak symmetry breaking constraints do not allow such choice, see \sect{EWSB};

\item
the RPV trilinear terms vanish
\beq
A_{ijk}=A'_{ijk}=A''_{ijk}=0 \, .
\eeq
\end{itemize}

We are now ready to study the phenomenology. We will first consider proton decay. 
Here there are two new issues compared to the RPC case. First, as we will see in the next section, 
an additional constraint must be taken into account in the unification analysis. 
Second, due to the huge sensitivity of proton decay to the exact value of $\lambda'' \approx 0$, 
new decay channels might contribute as well. After that we will systematically go through 
leptonic RPV consequences.

\subsection{Proton decay and unification constraints}

Although we will not dwell too much on the proton decay issue, some remarks are due. 
Unification of gauge couplings 
\cite{Dimopoulos:1981yj,Ibanez:1981yh,Einhorn:1981sx,Marciano:1981un} 
in the minimal renormalizable SU(5) model seems at odds with the experimental limits on proton decay if one assumes 
order TeV susy spectrum \cite{Murayama:2001ur}, albeit playing with the flavour 
structure of soft terms allows to solve the problem \cite{Bajc:2002bv,Bajc:2002pg}. Another logical possibility 
is simply to increase the susy scale. Nowadays, following ugly experimental facts and neglecting 
beautiful theoretical ideas, this is not a taboo anymore. In the usual RPC case it is enough to 
increase the susy scale to the multi-TeV region 
for low $\tan{\beta}$ in order to get the $d=5$ proton decay channel under control 
\cite{Hisano:2013exa,Bajc:2013dea}. The point is \cite{Bajc:2013dea} that by increasing the 
susy scale the color-triplet mass rises as well due to gauge coupling unification constraints. 
On the other side, this reduces the combination of the heavy gauge boson mass squared times the 
mass parameter of the adjoint. 
The gauge boson mass cannot be too low due to the 
$d=6$ proton decay channel, but in the RPC case the mass of the adjoint can practically 
take any value and so can be diminished at will. 

Once however R-parity conservation is abandoned, and the $\eta_i$ are of order one due to the doublet-triplet 
fine-tuning (\ref{triplet})-(\ref{doublet}), the adjoint mass cannot be too small because it mediates the 
type I + III seesaw mechanism for neutrino masses (see \eq{LN} and \sect{seesawGUT}), so it is bounded from below by 
around $10^{13}$ GeV. This means that we cannot increase the susy scale at will and 
so we may have some problem with proton decay constraints. 

Let us now estimate these scales. Denoting by $m_{\tilde f}$ the common sfermion mass (taken also 
as the matching scale between SM and MSSM), 
by $m_\lambda$ the common gaugino mass, by $\mu$ the Higgsino mass ($\mu_0\approx \mu_1$), 
by $M_T$ the heavy color triplet mass, by $M_V$ the heavy gauge boson mass (taken also as the 
matching scale between MSSM and SU(5)) and $M_{24}$ the common mass of the heavy adjoint fields 
(differences due to order one Clebsches are neglected), we can write the approximate relations 
\cite{Hisano:1992jj,Bajc:2013dea}
\bea
\left(\frac{M_T}{10^{15}\;{\rm GeV}}\right)^6&\approx&\left(\frac{\mu}{1\;{\rm TeV}}\right)^5 \, , \\
\left(\frac{M_V}{10^{16}\;{\rm GeV}}\right)^6\left(\frac{M_{24}}{10^{16}\;{\rm GeV}}\right)^3
&\approx&\left(\frac{1\;{\rm TeV}}{m_\lambda}\right)^2 \, , 
\eea
together with the experimental constraints from $d=5$ proton decay, $d=6$ proton 
decay, tree-level contribution to neutrino masses from the exchange of heavy 
mediators from the adjoint, and perturbativity, respectively:
\bea
\left(\frac{M_T}{10^{15}\;{\rm GeV}}\right)\left(\frac{m_{\tilde f}}{1\;{\rm TeV}}\right)^2
\left(\frac{1\;{\rm TeV}}{max(m_\lambda,\mu)}\right)\frac{1}{\tan{\beta}}&\gtrsim& 10^3 \, , \\
\left(\frac{M_V}{10^{16}\;{\rm GeV}}\right)&\gtrsim&1/3 \, , \\
\left(\frac{M_{24}}{10^{16}\;{\rm GeV}}\right)&\gtrsim&10^{-3} \, , \\
\left(\frac{M_T}{10^{15}\;{\rm GeV}}\right)\left(\frac{10^{16}\;{\rm GeV}}{M_V}\right)&\lesssim& 10 \, .
\eea
Hence, we immediately find an upper (lower) limit on the gaugino (sfermion) masses:
\bea
m_\lambda&\lesssim&10^6\;{\rm TeV}  \, , \\
m_{\tilde f}&\gtrsim&30\;{\rm TeV}\;\sqrt{\tan{\beta}}\left(\Theta(\mu-m_\lambda)+
\sqrt{m_\lambda/\mu}\;\Theta(m_\lambda-\mu)\right)  \, .
\eea
So, as an example, we can have at small $\tan{\beta}\approx2$ a common but relatively 
high-susy scale 
\beq
m_\lambda, m_{\tilde f}, \mu\approx 60\;{\rm TeV}  \, ,
\eeq
with
\bea
M_{24}&\approx& 10^{13.8}\;{\rm GeV} \, , \\
M_V, M_T&\approx&10^{16.5}\;{\rm GeV} \, .
\eea
In such a case the $d=5$ proton decay channel is the leading one and to be seen soon.

In all other solutions, the susy spectrum must be split with possibly light Higgsino 
and/or gauginos. It has to be stressed though that all we said so far is valid at most 
as an order of magnitude estimate, so that factors of few are possible. 

Finally, let us notice that we could also have proton decay contributions due to a slightly nonzero 
$\lambda''$. This would open up new decay channels, for example $B+L$ 
conserving \cite{Vissani:1995hp}, not present in the usual Weinberg classification (although $B+L$ 
conserving proton decay could be mediated by $d >6$ operators even in RPC GUTs, see for 
example \cite{Babu:2012vb}). However, due to the required 
smallness of $\lambda''$, nothing else except baryon 
number violating processes would change in our analysis.

\subsection{\label{EWSB}Electroweak symmetry breaking}

Our potential is (everything is real)\footnote{Analogously to the RPC case, 
the tree-level (color- and charge-preserving) minimum of the MSSM with RPV terms 
does not lead to spontaneous CP violation \cite{Masip:1998ui}.}
\beq
V=\frac{1}{2}
\bem
H_u^0 & \tilde{\nu}_\alpha
\eem
\bem
\mu_0^2+\mu_1^2+m_{H_u}^2 & -B_\beta \cr
-B_\alpha & \mu_\alpha\mu_\beta + m_{\alpha\beta}^2
\eem
\bem
H_u^0 \cr
\tilde{\nu}_\beta
\eem
+\frac{g^2+g'^2}{32}\left(\tilde\nu_\alpha^2-(H_u^0)^2\right)^2  \, ,
\eeq 
where $\alpha,\beta$ run from 0 to 1 (with $m^2_{01}=m^2_{10}$) 
and we consider the basis 
\bea
\langle H_u^0\rangle&=&v\sin{\beta} \, , \\
\langle \tilde{\nu}_0\rangle&\equiv& \langle H_d^0\rangle = v\cos{\beta} \, , \\
\langle \tilde{\nu}_1\rangle&=&0 \, .
\eea
The stationary equations give:
\bea
\label{mu0sq}
\mu_0^2&=&\frac{m_{00}^2-(m_{H_u}^2+\mu_1^2)\tan^2{\beta}}{\tan^2{\beta}-1}-\frac{\left(g^2+g'^2\right)v^2}{8} \, ,\\
\label{B0}
B_0&=&\frac{m_{00}^2-(m_{H_u}^2+\mu_1^2)}{\tan^2{\beta}-1}\tan{\beta} 
-\frac{\left(g^2+g'^2\right)v^2\tan{\beta}}{4\left(\tan^2{\beta}+1\right)} 
\, ,\\
\label{B1}
B_1&=&\frac{\mu_0\mu_1+m_{01}^2}{\tan{\beta}} \, .
\eea
This correctly reproduces the RPC case ($m_{00}=m_{H_d}$, $\mu_1=0$, $B_1=0$ and $m^2_{01}=0$). Notice that 
due to (\ref{B1}) we cannot take both $B_1$ {\it and} $m_{01}^2$ vanishing. This was the motivation 
for the assumptions (\ref{Bi}) and (\ref{m0i}).

By expanding $H_{u,d}^0=v_{u,d}+h_{u,d}^0$, 
the mass matrix of the neutral (real) scalars in the $(h_u^0, h_d^0, \tilde\nu_1)$ basis is found to be
\begin{multline}
\mathcal{M}_S^2 =\bem
(m_{00}^2-(m_{H_u}^2+\mu_1^2))\frac{1}{1-\tan^2\beta} &
(m_{00}^2-(m_{H_u}^2+\mu_1^2)) \frac{-\tan\beta}{1-\tan^2\beta} & 
\frac{m_{01}^2 + \mu_0 \mu_1}{\tan\beta} \\
(m_{00}^2-(m_{H_u}^2+\mu_1^2)) \frac{-\tan\beta}{1-\tan^2\beta} & 
(m_{00}^2-(m_{H_u}^2+\mu_1^2))\frac{\tan^2\beta}{1-\tan^2\beta} &
-\mu_0 \mu_1 - m_{01}^2 \\
\frac{m_{01}^2 + \mu_0 \mu_1}{\tan\beta} & 
- \mu_0 \mu_1 - m_{01}^2 & 
- \mu_1^2 - m_{11}^2
\eem \\ + \mathcal{O}(v^2)
\, , 
\end{multline}
where we also substituted the stationary conditions in \eqs{mu0sq}{B1} 
and we neglected $\mathcal{O}(v^2)$ terms. 
It is easy to see then, that the lightest eigenvalue (massless in the $v \to 0$ limit) 
is associated with the eigenvector $(\tan\beta,1,0)$. Hence, in the decoupling limit 
the light Higgs has no projections on the sneutrino direction. In the finite $v$ case the 
component of the light Higgs in the sneutrino direction is thus proportional to $v^2/m_{susy}^2$.

\subsection{Neutrino masses}
\label{numasses}

In this section we will see which constraints must be satisfied in order for neutrino masses to be 
in the right ballpark. In doing this, 
we will use the mass insertion approximation 
for the RPV bilinear couplings 
as e.g.~in \cite{Davidson:2000uc,Davidson:2000ne}. Although this is 
unjustified in the present context due to large RPV couplings, we assume that they give the 
right order of magnitude. The purpose of this calculation is not that of predicting neutrino masses but 
rather to check their consistency with experimental data. 
In particular, we will estimate (in order of importance): 
the tree-level seesaw contribution from RPV interactions,
the leading one-loop RPV corrections and 
the type I + III seesaw contribution from GUT-scale mediators. 
Let us now discuss in turn the various cases. 

\subsubsection{Tree-level seesaw from RPV interactions}

This is the most important contribution. 
By neglecting the typically much smaller GUT-scale induced 
type I + III seesaw contribution (to be discuss in \sect{seesawGUT}), 
the only non-vanishing element of the neutrino mass matrix is
\beq
\label{m11}
m_{11}=-\frac{\mu_1^2 v^2\cos^2{\beta}}{4(\mu_0^2+\mu_1^2)}\left(\frac{g'^2}{M_1}+\frac{g^2}{M_2}\right) \, ,
\eeq
where we expanded in $v/M_{1,2}$, while keeping $\mu_1 / \mu_0$ of order one \cite{Bisset:1998bt} (see also Appendix \ref{pertdiag}). 
\eq{m11} can be made small, for our choice of parameters, only assuming a very strong cancellation
\beq
\label{gaugini}
\cos^2{\beta}\left(\frac{g'^2}{M_1}+\frac{g^2}{M_2}\right)\lesssim 10^{-13}\;{\rm GeV}^{-1} \, , 
\eeq
i.e.~having gaugino masses with opposite sign and fine-tuned ratio. 
This is possible since we did not assume any specific boundary condition 
on the soft terms (e.g.~gaugino masses unification). In \sect{conclusions} we will 
shortly comment on possible mechanisms of susy breaking which might yield to relations 
close to \eq{gaugini}.

Notice that 
the combination of gaugino masses in \eq{gaugini} is proportional to the photino 
mass parameter, $m_{\tilde{\gamma}} = M_1 c_W^2 + M_2 s_W^2$, and 
that the exact determinant of the generalized neutralino mass matrix in \eq{LN} 
(after restricting to the nontrivial rank-5 subspace and for $\eta_\alpha=0$) is still proportional to $m_{\tilde{\gamma}}$.  
Though $m_{\tilde{\gamma}} \rightarrow 0$ can be effectively used to suppress large tree-level neutrino masses, this limit does not seem to be associated with any new symmetry of the Lagrangian. In fact, already at one loop this fine-tuning is not enough anymore, 
since the rank of the neutrino mass matrix will change as we will see in the next subsection.

\subsubsection{One-loop contributions from RPV couplings}

\begin{figure}[htb]
\begin{center}
\includegraphics[width=15.cm]{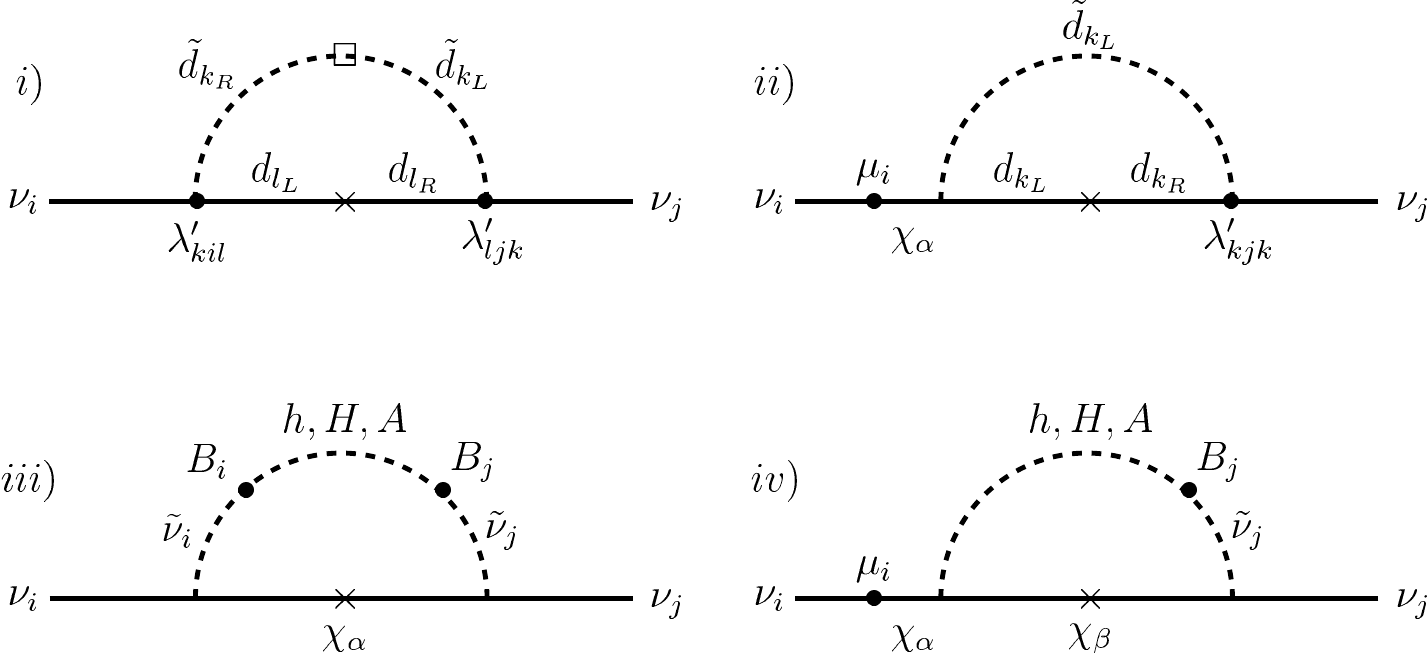}
\caption{\label{numass1loop} 
The dominant one-loop contributions to the neutrino mass matrix. 
The white square denotes a LR insertion in the squark mass matrix, while the cross stands for a mass insertion 
on the internal quark or neutralino propagator. The blob is associated with the source of R-parity violation. 
}
\end{center}
\end{figure}

The most relevant diagrams for the RPV one-loop corrections to the neutrino mass matrix \cite{Davidson:2000uc,Davidson:2000ne,Hirsch:2000ef,Diaz:2003as,Grossman:2003gq,Hirsch:2004he} are shown in \fig{numass1loop}.
Let us now estimate their size.
\begin{enumerate}[$i)$]

\item 
A standard computation gives
\beq
\label{diagi}
\delta m_{ij}\approx\frac{3}{16\pi^2}\sum_{k,l}\lambda'_{kil}\lambda'_{ljk}\frac{m_{d_k}m_{d_l}}{\tilde m^2_{d_k}}(A-\mu\tan{\beta}) \, .
\eeq
Taking $\tan{\beta}=10$, $\mu=-1$ TeV, $A=10$ TeV, $\tilde m_{d_i}=30$ TeV, $m_{d_3} = 4.2$ GeV and the fitted values of $\lambda'$ 
in \eqs{lamp3num}{lamp1num}, 
we get the elements of the lower right $2\times2$ block of the order of $100$ eV, definitely too large. 
One can suppress these contributions by another cancellation between $A$ and $\mu\tan{\beta}$ 
and/or by increasing the sfermion masses. Similar contributions come also from two $\lambda$s without 
the color factor and with sleptons running in the loop.

\item
Here the diagrams include the external neutrino mixing with both bino and 
wino through Higgsino; after summing all contributions and choosing a 
renormalization scheme such that the wino-neutrino mixing is canceled at the one-loop level 
\cite{Davidson:2000ne}, one gets various contributions each of the order of
\beq
\label{diagii}
\delta m_{ij}\approx\frac{3}{16\pi^2}\frac{g'^2v\cos{\beta}}{2\mu M_2}
\sum_{k}m_{d_k}\left(\mu_i\lambda'_{kjk}+\mu_j\lambda'_{kik}\right) \, .
\eeq
A more detailed calculation \cite{Davidson:2000ne} gives an exact cancellation in the 
degenerate down squark case ($\tilde m_{d_L}^2=\tilde m_{d_R}^2$). Similar 
diagrams with $\lambda'\to\lambda$ and sleptons in the loop require degenerate 
sleptons ($\tilde m_{\tau_L}^2=\tilde m_{\tau_R}^2$) for an exact cancellation. 

\item $+ \; $  $iv)$
These contributions can be written as \cite{Davidson:2000uc,Grossman:2003gq}
\beq
\label{34}
\delta m_{11}\approx \frac{g^2}{64\pi^2\cos^2{\beta}}\frac{B_1^2m_W^2}{m_{\tilde f}^6}M_2+ 
\frac{g^2}{64\pi^2\cos{\beta}}\frac{B_1\mu_1m_W^2}{m_{\tilde f}^4} \, , 
\eeq
where we assumed $m_W\ll M_2\ll m_{\tilde f}\approx m_{H_{u,d}}$. 
Notice the $m_W^2/m_{\tilde f}^2$ suppression in \eq{34}, which is a remnant 
of an exact cancellation of the loop functions in the decoupling limit \cite{Grossman:2003gq}.
These contributions are in the same 
direction as the fine-tuned tree-level one. So all one needs is  doing just a slightly different fine-tuning. 

\end{enumerate}

\subsubsection{Seesaw from GUT-scale mediators}
\label{seesawGUT}

For completeness, we estimate the rank-1 type I + III seesaw contribution from GUT-scale mediators in \eq{LN} 
in the limit of no RPV mixing. 
This yields one non-vanishing neutrino mass eigenvalue
\begin{equation}
\label{mnuapprox}
m_{\nu}= \eta^2v_u^2/M_{\text{seesaw}} \, , 
\end{equation}
where $\eta=\sqrt{\eta_0^2+\eta_k^2}$ and $\eta = \mathcal{O}(1)$ in order to achieve the doublet-triplet splitting (cf.~\sect{RPVSU5}).  
Notice that, since $M_{\text{seesaw}}$ could be as large as $M_{GUT} \gtrsim 10^{16}$ GeV, this contribution to neutrino masses can be made subleading. 

In conclusion, neutrino masses can be (admittedly barely) under control assuming a strong fine-tuning 
among wino and bino mass parameters (\ref{gaugini}) to suppress the tree-level contribution, heavy 
sfermions or small left-right sfermion mixings to suppress (\ref{diagi}), 
an approximate degeneracy in the sfermion spectrum to suppress the one-loop contribution (\ref{diagii}) 
and $M_{\text{seesaw}} \approx M_{GUT}$.

\subsection{Modifications of SM couplings to leptons}
\label{SMcouplep}

The mixing between leptons and higgsinos/gauginos is also constrained 
by the measurement of the SM couplings to 
the lightest lepton mass eigenstates $\hat{e}_{1,2,3}$ and $\hat{\nu}_{1,2,3}$.
The relevant couplings to be considered here are: 
$Z \hat{e}_i \hat{e}_j$ (precision measurement at the $Z$ pole and lepton flavour violating charged lepton decays),  
$Z \hat{\nu}_i \hat{\nu}_j$ (invisible $Z$ width) and 
$W \hat{e}_i \hat{\nu}_j$ (charged lepton universality).

Assuming real parameters and denoting the deviation from a SM coupling $g_{\rm{SM}}$ as $\delta g_{\rm{SM}}$, 
the modified SM couplings to leptons are found to be (see also \cite{Bisset:1998bt,Bisset:1998hy,Nowakowski:1995dx}): 
\begin{itemize}
\item $Z \hat{e}_i \hat{e}_j$ couplings: 
\begin{align}
\delta g_L^{ij} &=  U_L^{i+2,1} U_L^{j+2,1} \, , \\
\delta g_R^{ij} &= 2 U_R^{i+2,1} U_R^{j+2,1} + U_R^{i+2,2} U_R^{j+2,2} \, , 
\end{align} 
where $U_{L,R}$ are the bi-unitary matrices which diagonalize the generalized chargino mass matrix (cf.~\app{pertdiag}), 
while $i$ and $j$ run over the three lightest eigenvalues. 
In particular, in the susy-decoupling limit considered in \app{pertdiag} 
we get:
\begin{align}
|U_L^{31}| & = \left| \frac{g v_d \mu_1}{\sqrt{2} \mu M_2} \right| = \mathcal{O}\left(m_W/M_2\right) \, , \\
|U_R^{31}| & = \left| \frac{\mu_1 m_1}{\mu^2} \frac{g v_u}{\sqrt{2} M_2} \right| = \mathcal{O}\left(m_1 m_W / (\mu M_2)\right)  \, , \\
|U_R^{32}| & = \left| \frac{\mu_1 m_1}{\mu^2} \right| = \mathcal{O}\left(m_1/\mu\right) \, , 
\end{align}  
and the modified couplings of the $Z$ boson to charged leptons (electrons) are hence $\delta g_L^{11} = \mathcal{O}\left(m_W^2/M_2^2\right)$ 
and $\delta g_R^{11} = \mathcal{O}\left(m_1^2/\mu^2\right)$. 

The constraints from the $Z$-pole observables are typically given in terms of $\delta g_{V,A} = \tfrac{1}{2}(\delta g_L \pm \delta g_R)$ 
and are at most at the $0.07 \%$ level for the flavour diagonal case 
\cite{Frugiuele:2011mh,Riva:2012hz,Agashe:2014kda}. 
On the other hand, the bounds on the flavour violating couplings are less strict, with the only exception of those 
coming from the measurement of $\mu\to eee^c$, which sets 
$\delta g_{V,A}^{12} \lesssim 10^{-6}$ \cite{Bisset:1998hy,Davidson:2012wn}. 
The latter bound is evaded by our specific flavour orientation of the $\mu_i$ vector, 
e.g.~$\mu_i \propto \delta_{1i}$. 

Hence, all the relevant bounds due to the modification of the $Z$ boson couplings to charged leptons 
are satisfied by $M_2 \gtrsim 5$ TeV and $\mu_i \propto \delta_{1i}$.

\item $Z \hat{\nu}_i \hat{\nu}_j$ couplings: 
\beq
\delta g_{V,A}^{ij} = - U_0^{i+4,1} U_0^{j+4,1} - U_0^{i+4,2} U_0^{j+4,2} - 2 U_0^{i+4,3} U_0^{j+4,3} \, ,
\eeq
where $U_{0}$ is the unitary matrix which diagonalizes the generalized neutralino mass matrix (cf.~\app{pertdiag}), 
while $i$ and $j$ run over the three lightest eigenvalues. 

At the leading order in the expansion of \app{pertdiag}, we find 
\begin{align}
|U_0^{51}| & = \left| \frac{g' v_d \mu_1}{2 \mu M_1} \right| = \mathcal{O}\left(m_W/M_1\right) \, , \\
|U_0^{52}| & = \left| \frac{g v_d \mu_1}{2 \mu M_2} \right| = \mathcal{O}\left(m_W/M_2\right)  \, .
%|U_0^{i3}| & \approx 0 \, . 
\end{align}  
For $\mu > m_Z$, the typical signature is the reduction of the invisible width of the $Z$ boson. 
However, even for moderate (non-decoupled) values of $M_{1,2}$, the inferred bound on $\mu_1$ is very mild \cite{Bisset:1998hy}. 
 
\item $W \hat{e}_i \hat{\nu}_j$ couplings: 

Defining the current eigenstate matrices 
\begin{align}
T^L &=
\bem
0 & \sqrt{2} & 0 & 0 \\
0 & 0 & 0 & 1_{4 \times 4} 
\eem \, , \\
T^R &=
\bem
0 & -\sqrt{2} & 0 & 0 \\
0 & 0 & 1 & 0_{4 \times 4} 
\eem \, , 
\end{align}
the modified SM couplings read
\begin{align}
\delta \tilde{g}_L^{ij} &=  (U_L^\dag T^L U_0)^{ij} \, , \\
\delta \tilde{g}_R^{ij} &= (U_R^\dag T^R U_0)^{ij} \, , 
\end{align}  
where $i$ and $j$ run over the three lightest eigenvalues.
Charged lepton universality in charged current processes, such as the decay of pions and leptons,  
is experimentally verified at the $0.2 \%$ level \cite{Loinaz:2004qc}. This typically yields less stringent bounds than those 
derived from $Z$ couplings \cite{Bisset:1998hy}. 

\end{itemize}

Summarizing, the couplings of the $Z$ and $W$ bosons to the three lightest lepton mass eigenstates can be easily made compatible 
with the SM values by a moderate decoupling of gaugino masses (say $M_{1,2} \gtrsim 5$ TeV) and for $\mu_i \propto \delta_{i1}$. 
This was indeed to be expected, since in the 
gaugino decoupling limit we are mixing only representations with the same gauge quantum numbers (GIM-like mechanism), and hence gauge couplings have to be SM-like.

\subsection{Other lepton number violating processes}
\label{LNvp}

On top of neutrino masses there are also other lepton number violating effects which are worth to 
be discussed. First of all, LHC can produce via a Drell-Yan process a pair of winos 
which can subsequently decay through lepton number 
violating couplings into same-sign dileptons \cite{Keung:1983uu} and 4 jets with no missing energy 
(ideally, a background-free process):
\beq
pp\to W^{*\pm}\to\tilde W^\pm\tilde W^0\to(e^\pm Z)(e^\pm W^\mp)\to(e^\pm jj)(e^\pm jj) \, . 
\eeq

This is completely analogous the the production and decay of a  light weak triplet fermion pair 
from type III seesaw \cite{Bajc:2006ia,Bajc:2007zf,Arhrib:2009mz,DiLuzio:2013dda}. 
Since winos are unstable the cross section $\sigma(pp\to\tilde W^\pm\tilde W^0)$ gets multiplied with 
an approximate factor
\beq
\int_{E_{min}^2}^{E_{max}^2}\frac{M_2\Gamma dp^2}{(p^2-M_2^2)^2+M_2^2\Gamma_{TOT}^2} \, , 
\eeq
for each wino.
For $E_{min}^2\ll M_2^2\ll E_{max}^2$ the integral can be approximated by the branching fraction of the decay channel. 
This is what happens in the usual MSSM with light $M_2$ and small RPV couplings. 

However, since in our case winos are typically much heavier than the electroweak scale 
($M_2 \gtrsim 5$ TeV from the modified $Z$ couplings -- see \sect{SMcouplep}), 
we should replace (very roughly)
\begin{multline}
\label{BRBR}
BR(\tilde W^\pm\to e^\pm Z)BR(\tilde W^0\to e^\pm W^\mp)\longrightarrow \\ 
\left(\frac{E_{max}}{M_2}\right)^4
\left(\frac{\Gamma(\tilde W^\pm\to e^\pm Z)}{M_2}\right)\left(\frac{\Gamma(\tilde W^0\to e^\pm W^\mp)}{M_2}\right) \, .
\end{multline}
This is small due to the $(m_W/M_2)^2$ suppression of the $\Gamma$ (see \eq{mixwhd}) and eventually because 
$E_{max} < M_2$.
Hence, in spite of the fact that the RPV coupling $\mu_1/\mu_0$ is much larger than in the usual 
case, this lepton number violating process will not be easily accessible at LHC because the ratio 
$m_W/M_2\lesssim1/50$ is too small, giving for \eq{BRBR} a suppression of $\approx 10^{-7}$.

The next lepton number violating process we consider is neutrinoless double $\beta$ decay. 
Following \cite{Barbier:2004ez} the limits on the trilinear RPV couplings are ($k=1,2,3$)
\bea
|\lambda'_{111}|^2\left(\frac{m_W}{m_{\tilde f}}\right)^4\left(\frac{m_W}{m_{\lambda}}\right)&\lesssim& 10^{-8} \, , \\
|\lambda'_{11k}\lambda'_{k11}|\left(\frac{m_W}{m_{\tilde f}}\right)^4\left(\frac{A-\mu\tan{\beta}}{m_W}\right)&\lesssim& 10^{-(6\div8)} \, ,
\eea
which are easily satisfied in our case, even for relatively low super-partner masses. On the other hand, the 
parameter $\mu_1/\mu_0$ contributes to the process only through the light neutrino masses, 
whose suppression has been already discussed in \sect{numasses}.  

Finally, other potentially relevant lepton number violating processes like e.g.~$\mu^+\to e^-$ conversion in nuclei, 
$K^+ \to \mu^+ \mu^+ \pi^-$ or $\bar{\nu}_e$ emission from the Sun, 
do not bring any really important constraint on the model parameters 
since the experimental limits on the branching ratios are still too weak. 

\subsection{Lepton flavour violation}
\label{LFVsect}

In this section we analyse in more detail lepton flavour violating processes 
like $\mu \to e$ conversion in nuclei, $\mu\to eee^c$ and 
$\mu\to e\gamma$ (other processes involving the $\tau$ lepton 
are worse measured and their bounds can be easily evaded).
At leading order ($\epsilon^0)$ in $\epsilon=\mathcal{O}(m_W / M_2, m_1 / m_W)\lesssim 10^{-2}$ there is no mixing 
between generations, i.e.~the electron mass eigenstate mixes just with Higgsino, while the muon does 
not mix at all ($\mu_2=0$), see Appendix \ref{pertdiag}. In other words, at order $\epsilon^0$ and tree level 
the $\lambda$ and $\lambda'$ couplings are already in the mass eigenbasis. 
In particular, all the lepton flavour changing amplitudes involving electrons
vanish at order $\epsilon^0$. Following for example 
the computation and notation of \cite{Kim:1997rr} for $\mu \to e$ conversion and 
\cite{deGouvea:2000cf} for the other two processes, we can summarize the results as 
follows ($\lambda$ and $\lambda'$ corresponding to the values determined in \sect{trilinearRPV}):
\begin{itemize}
\item 
$\mu \to e$ conversion: 
the coefficients in front of the possible operators of the type 
$\bar e \mu\bar qq$ are at tree order 
\bea
\label{mue1}
A^d\sim+\sum_{k=1}^3\frac{\lambda'_{11k}\lambda'_{12k}}{m_{\tilde Q_k}^2}\to0 \, , &&
A^u\sim-\sum_{k=1}^3\frac{\lambda'_{k11}\lambda'_{k21}}{m_{\tilde d_k}^2}\to0  \, , \\
S^{d,1}\sim-2\sum_{k=1}^3\frac{\lambda'_{1k1}\lambda_{k12}}{m_{\tilde Q_k}^2}\to0 \, , &&
S^{d,2}\sim-2\sum_{k=1}^3\frac{\lambda'_{1k1}\lambda_{k21}}{m_{\tilde Q_k}^2}\to0 \, .
\eea

\item 
$\mu\to eee^c$: 
the coefficients in front of the possible operators of the type 
$\bar e \mu\bar ee$ are at tree order 
\beq
B^L\sim-\sum_{k=1}^3\frac{\lambda_{k11}\lambda_{k21}}{2 m_{\tilde L_k}^2}\to0\, , \;\;\;\;\;\;
B^R\sim-\sum_{k=1}^3\frac{\lambda_{k11}\lambda_{k12}}{2 m_{\tilde L_k}^2}\to0 \, .
\eeq

\item
$\mu\to e\gamma$: 
the coefficients in front of the possible operators are at one-loop order
\begin{align}
A_2^R&\sim\frac{1}{16\pi^2}\frac{1}{12}\sum_{j,k=1}^3\left(-2\frac{\lambda_{1kj}\lambda_{2kj}}{m_{\tilde L_k}^2}
+\frac{\lambda_{1kj}\lambda_{2kj}}{m_{\tilde e_k}^2}
-3\frac{\lambda'_{k1j}\lambda'_{k2j}}{m_{\tilde d_k}^2}\right)\to0 \, , \\
\label{muegamma2}
A_2^L&\sim\frac{1}{16\pi^2}\frac{1}{12}\sum_{j,k=1}^3\left(-2\frac{\lambda_{kj1}\lambda_{kj2}}{m_{\tilde L_k}^2}
+\frac{\lambda_{jk1}\lambda_{jk2}}{m_{\tilde L_k}^2}\right)\to0 \, .
\end{align}

\end{itemize}
Next we want to check what happens beyond the leading order.
Without doing a full calculation for the order $\epsilon$ or at higher loops, we can consider the following:
\begin{enumerate}

\item
Either $\epsilon$ or an extra loop factor contribute with a suppression factor of at least $10^{-2}$;

\item
Although $L_1$ violation is in principle order one, $L_2$ violation is of order $10^{-1}$ (cf.~discussion below \eq{lamp1num});

\item
The propagator gets a suppression $(m_W/m_{\tilde f})^2$ compared to the Fermi constant $G_F$

\end{enumerate}
Putting all this together, we schematically find for the generic coefficient ${\mathcal A}$ 
in \eqs{mue1}{muegamma2} relative to the different processes:

\begin{itemize}

\item
$\mu \to e$ conversion: comparing  theoretical expectations \cite{Kitano:2002mt} 
with the experimental constraint on Titanium \cite{Dohmen:1993mp}
\beq
m_W^2{\mathcal A}_{\mu-e}\sim 10^{-2}10^{-1}\left(\frac{m_W}{m_{\tilde f}}\right)^2\lesssim 10^{-7} \, ,
\eeq
which can be satisfied for sfermion masses of order $10$ TeV or more. 

\item 
$\mu\to eee^c$: similar estimates give (see \cite{Agashe:2014kda} for experimental 
bounds)
\beq
m_W^2{\mathcal A}_{\mu\to3e}\sim 10^{-2}10^{-1}\left(\frac{m_W}{m_{\tilde f}}\right)^2\lesssim 10^{-6} \, ,
\eeq
again easily satisfied for $\tilde m_f\gtrsim 3$ TeV.

\item 
$\mu\to e\gamma$: following again \cite{Agashe:2014kda} we find (notice that here we started already at one-loop) 
\beq
m_W^2{\mathcal A}_{\mu\to e\gamma}\sim (10^{-2})^210^{-1}\left(\frac{m_W}{m_{\tilde f}}\right)^2\lesssim 10^{-6} \, , 
\eeq
which is evaded already for $m_{\tilde f}\gtrsim 300$ GeV.

\end{itemize}

\subsection{Gravitino dark matter}

In the presence of sizeable RPV interactions there are no long-lived states in the MSSM 
spectrum, so the only DM candidate is a slowly decaying gravitino. 
For $m_{3/2} < m_Z$ the main decay channel of the gravitino is \cite{Takayama:2000uz}
\beq
\label{GammaG}
\Gamma (\tilde{G} \rightarrow \gamma \nu) = \frac{1}{32 \pi} |U_{\tilde{\gamma} \nu}|^2 \frac{m^3_{3/2}}{M_P^2} \, , 
\eeq
where $U_{\tilde{\gamma} \nu} = c_W U_{\tilde{B} \nu} + s_W U_{\tilde{W} \nu}$ is the photino-neutrino mixing 
and $M_P = 2.4 \times 10^{18}$ GeV is the reduced Planck mass. 
From \eq{nu2eps} we read 
\beq
\label{Ugammanu}
|U_{\tilde{\gamma} \nu}|^2 =  \frac{\pi \alpha_{\rm{em}} v_d^2 \mu_1^2}{\mu^2} \left( \frac{1}{M_1} - \frac{1}{M_2} \right)^2 
\approx 10^{-7} \left( \frac{10}{\tan\beta} \right)^2 \left( \frac{10 \ \text{TeV}}{M_{1}} \right)^2 
\, ,  
\eeq
where $\tan\beta \gg 1$ and we already considered the fine-tuning in \eq{gaugini} in order to suppress neutrino masses. 
This has to be compared with the standard case where the smallness of neutrino masses is due to a tiny 
mixing with gauginos, yielding \cite{Takayama:2000uz} 
\beq
|U_{\tilde{\gamma} \nu}|^2_{\rm{stand}} = \mathcal{O} \left( \frac{m_\nu}{M_{1}} \right) \, ,  
\eeq
or, equivalently
\beq
|U_{\tilde{\gamma} \nu}|^2 / |U_{\tilde{\gamma} \nu}|^2_{\rm{stand}} \approx 10^6 \left( \frac{10}{\tan\beta} \right)^2  \left( \frac{10 \ \rm{TeV}}{M_{1}} \right) \, .  
\eeq
Hence, in our scenario, where neutrino masses and $U_{\tilde\gamma\nu}$ mixing are 
decoupled, the gravitino decays a factor $\approx 10^6$ faster than 
in the standard RPV case and so we have to check whether it can still be a good DM candidate. 

As a first check let us compare its lifetime with the age of the Universe $\tau_U \approx 4.3 \times 10^{17}$ s. 
From \eqs{GammaG}{Ugammanu} we obtain  
\beq
\label{lifetime32}
\tau_{3/2} \approx 3.8 \times 10^{18} \ \text{s} \left( \frac{\tan\beta}{10} \right)^2 \left( \frac{M_{1}}{10 \ \text{TeV}} \right)^2 \left( \frac{10 \ \text{GeV}}{m_{3/2}} \right)^3 \, ,
\eeq
which is safe, as long as $m_{3/2} \lesssim 10$ GeV (for $M_{1} \approx 10$ TeV and $\tan\beta \approx 10$).

The decay of the gravitino is expected to leave an imprint on the 
extragalactic diffuse high-energy photon background in the form 
of a monochromatic line centred at $m_{3/2} / 2$. 
This is because $m_{3/2}$ is very light, contrary to what happens with multi-TeV 
gravitino masses where a continuum signal in the spectrum is expected, 
see for example \cite{Bajc:2010qj}. 
The photon number flux, $F_{\gamma}^{\rm{max}}$, at the peak of the maximum photon energy $E_\gamma = m_{3/2}/2$, 
is estimated to be \cite{Takayama:2000uz}
\beq
F_{\gamma}^{\rm{max}} \approx 10^{-5} \ \text{cm}^{-2} \ \text{sr}^{-1} \ \text{s}^{-1} \left( \frac{m_{3/2}}{10 \ \rm{MeV}} \right)^2 
\left( \frac{\Omega_{3/2} h^2}{0.12} \right) \left( \frac{10}{\tan\beta} \right)^2 \left( \frac{10 \ \rm{TeV}}{M_{1}} \right)^2 \, , 
\eeq 
which is compatible with the bounds coming from diffuse X- and gamma-ray fluxes 
\cite{UweOberlack,Abdo:2010nz,Essig:2013goa}, 
as long as $m_{3/2} \lesssim 10$ MeV 
(for $M_{1} \approx 10$ TeV and $\tan\beta \approx 10$). 
The latter values correspond to a lifetime $\tau_{3/2} > 10^{27\div 28}$ s, 
which is indeed the typically constraint for decaying DM into photons \cite{Herder:2009im}. 
Notice, also, that there are no observational constraints (from Big Bang Nucleosynthesis or CMB) 
on the decay of the next-to-lightest supersymmetric particle, due to its fast decay via large RPV interaction.

The last point we want to address is a possible constraint related to the reheating temperature. 
Assuming thermal production in the early Universe, the gravitino relic density is constrained by 
(see e.g.~\cite{Pradler:2006qh,Rychkov:2007uq,Olechowski:2009bd,Arcadi:2011yw})
\beq
\Omega_{3/2} h^2 \gtrsim 0.12 \left( \frac{T_{RH}}{300 \ \text{GeV}} \right) \left( \frac{10 \ \text{MeV}}{m_{3/2}} \right) 
\left( \frac{M_{2}}{30 \ \text{TeV}} \right)^2 \, , 
\eeq 
where approximate equality holds when the gluino contribution can be neglected. 
Notice that for $m_{3/2} \lesssim 10$ MeV and $M_{2} \approx 30$ TeV 
($M_1\approx -M_2g'^2/g^2 \approx 9$ TeV), 
the reheating temperature can still be above the electroweak phase transition. 
On the other hand, gravitino masses lighter than already 1 MeV (or, equivalently, too large gaugino masses) would imply a reheating 
temperature well below the electroweak phase transition, 
which is difficult to reconcile with an high-energy mechanism of baryogenesis.\footnote{In our setup 
baryon number is effectively preserved below the GUT scale, therefore we are only left with the possibility 
of generating a lepton asymmetry above the electroweak phase transition and get it converted 
into a baryon one through sphalerons effects.} 
From this point of view, a gravitino mass close to the upper limit 
of $10$ MeV (compatible with the measured photon fluxes) is theoretically 
favourable. This is, of course, also the most interesting region for a possible experimental discovery.

\section{Discussion and conclusions}
\label{conclusions}

Among grand unified theories only renormalizable SO(10) 
\cite{Clark:1982ai,Aulakh:1982sw,Aulakh:2003kg} is able to derive exact 
R-parity conservation \cite{Mohapatra:1986su,Font:1989ai,Martin:1992mq} at low energies 
\cite{Aulakh:1997ba,Aulakh:1998nn,Aulakh:1999cd}, while there 
is no reason to assume it in SU(5). There are of course strong phenomenological constraints 
that make especially the baryon number violating couplings practically zero. In this work 
we tried to see if the remaining R-parity violating interactions in the minimal renormalizable SU(5) 
can be of any utility for the down quark vs.~charged lepton mass problem of the original setup. 
The outcome of our analysis is positive: these couplings are able to reproduce the SM fermion 
masses and so avoid large susy breaking threshold corrections which  
would make our vacuum metastable \cite{Casas:1995pd}. 
The prize to pay are three classes of fine-tuning: $i)$ a generalized doublet-triplet splitting 
(cf.~\eqs{mualpha}{Malpha}), $ii)$ the vanishing of the baryonic RPV 
couplings $\lambda''$ in \eq{lambdaseq0} and $iii)$ 
the suppression of neutrino masses in \eq{gaugini}.

Is relation (\ref{gaugini}) between gaugino masses a prediction of the 
theory? Since it gets corrections at higher loops, \eq{34} being 
the dominant one, the question is thus: how exactly must $M_1/g'^2=-M_2/g^2$ hold? Let us see what 
we need for this relation to be for example $10\%$ exact, i.e.~suppose
\beq
\label{M12}
\frac{M_1}{g'^2}=-\frac{M_2}{g^2}(1\pm0.1) \, .
\eeq
This is equivalent to say that the loop contribution is at most $10\%$ of the  
{\it non-fine-tuned} value in \eq{m11}, i.e.
\beq
\delta m_{11}\lesssim \frac{1}{10}\times\frac{\mu_1^2v^2\cos^2{\beta}}{4(\mu_0^2+\mu_1^2)}\frac{g^2}{M_2} \, .
\eeq
In usual perturbation theory 
$\delta m/m$ is loop suppressed, so small, provided the same couplings as at tree order are used. 
But in our case we have more like a Coleman-Weinberg situation \cite{Coleman:1973jx}, where new couplings 
not present at tree level, in our case $B_1$, start contributing. So there is 
no limitation from perturbation theory and at least in principle loops could dominate over tree-level 
contributions. Is this what happens here? According to (\ref{34}), and assuming a split susy spectrum 
$\mu_1\sim M_2\ll m_{\tilde f}\sim\sqrt{|B_1|}$ we find that very roughly the $10\%$ 
correlation between bino and wino mass (\ref{M12}) is valid if
\beq
\label{M2}
M_2\lesssim 10\cos^2{\beta}\;m_{\tilde f} \, .
\eeq
For larger $M_2$, there is still a strong correlation between $M_1$ and $M_2$, but other parameters 
get involved too, so it is harder to make a definite statement of what to look for.  
But if \eq{M2} is valid, 
the apparently weak point of the neutrino mass becomes a strong one, and the theory is falsifiable through 
a future experimental check of \eq{M12}.

Suppose now that $M_2$ satisfies \eq{M2}. 
Is there any obvious theoretical reason why would \eq{M12} hold? 
In other words, can one find a susy breaking 
and mediation mechanism which leads to it at least at the one-loop level? A natural candidate would be gauge mediation. 
The change in sign of the bino mass compared to wino mass can be obtained only by a combination 
of gauge messengers (which contribute negatively) with chiral messengers  (which contribute positively). 
A naive simple computation shows that if an SU(5) adjoint breaks susy like for example in 
\cite{Bajc:2008vk,Bajc:2012sm}, one needs \cite{Giudice:1997ni}
\beq
1=\frac{(M_1/g_1^2)}{(-3/5)(M_2/g_2^2)}=\frac{(\Delta b_{chiral}-10)}{(-3/5)(\Delta b_{chiral} -6)} \, ,
\eeq
where we assumed that chiral superfields contribute in SU(5) multiplets. The change of the SU(5) beta function equals 
$\Delta b_{chiral}=17/2$ on the threshold. 
A half-integer $\Delta b_{chiral}$ seems impossible to obtain: a 
complex representation needs always to come in pairs to be vector-like and satisfy anomaly constraints, 
while real representations have an integer Dynkin index. Evading this conclusion needs more sophisticated 
scenarios. However, if (\ref{M12}) is relaxed a bit (by $M_{1,2}\gtrsim 10 \ m_{\tilde f}$ and/or large $\tan{\beta}$), 
then we can get with an integer $\Delta b_{chiral}$ (for example 8 or 9) opposite sign bino and wino masses.

Another possibility is to consider gravity mediation. From \cite{Martin:2009ad}\footnote{We thank Ilia 
Gogoladze for pointing out this possibility.} we see that relation (\ref{M12}) is obtained for example in 
SO(10) if a $210$ is coupled to gauge field strength bilinears and its parity odd Pati-Salam singlet gets 
a non-zero F-term. Although amusing, it is unclear what this means in the context of our renormalizable 
SU(5) model.

On top of \eq{M12}, the other prediction of the model is a gravitino dark matter candidate lighter than approximately $10$ MeV, 
preferably closer to the upper limit in order to be reconcilable with baryogenesis. 
A gravitino mass in the region favoured by baryogenesis is also 
the most interesting one from an experimental point of view. 
The main signature being a monochromatic line 
in the diffuse extragalactic photon background picked around $5$ MeV.

In this work we only used the RPV mixing effects to correct the wrong SU(5) mass 
relations. In practice, however, the solution to this problem could arise from different sources, 
partially from susy threshold corrections and partially from RPV mixings, thus modifying 
the numerical values of the RPV parameters here considered. 
Also, the ad-hoc assumption of setting to zero those couplings 
that make the wrong mass relations worse, is not really needed, although a generic situation might be forbidden 
by data. 

Although the model is a bit stretched and many tunings of parameters are needed, the 
phenomenology itself seems interesting: the electron mass eigenstate (or other leptons as 
well in a more general framework) may not be what we usually think of, but rather an order 
half-electron and half-Higgsino flavour state.

\section*{Acknowledgments}
We are deeply indebted with Sacha Davidson for several valuable discussions, clarification regarding 
her work and correspondence. We thank Marco Nardecchia for collaboration in the early stages of this 
project, Giorgio Arcadi, Ilia Gogoladze, Miha Nemev\v sek and Gabrijela Zaharijas for discussions, 
Stephane Lavignac, Goran Senjanovi\'c and Vasja Susi\v c for reading the manuscript and useful comments. 
We thank Kaladi Babu for pointing out an error in a previous version of this paper. 
The work of B.B.~has been supported by the Slovenian Research Agency. 
B.B.~would like to thank CETUP* (Center for Theoretical Underground Physics and Related Areas), supported by the 
US Department of Energy under Grant No.~DE-SC0010137 and by the US National Science Foundation under Grant No. 
PHY-1342611, for its hospitality and partial support during the 2013 Summer Program. B.B.~acknowledges the hospitality 
and support from the NORDITA scientific program ÒNews in Neutrino PhysicsÓ, April 7-May 2, 2014, during which part of 
this study was performed. The work of L.D.L.~is supported by the Marie Curie CIG program, project number PCIG13-GA-2013-618439. 
L.D.L.~is grateful to the theoretical phyisics group of the Jo\v{z}ef Stefan Institute 
for the warm hospitality and support during the development of this project. 

\appendix

\section{Perturbative diagonalization}
\label{pertdiag}

Let us write the diagonalization of the generalized chargino and neutralino mass matrices, in \eq{C} and \eq{LN} respectively, as: 
\begin{align}
U_L^\dag \mathcal{M}_C U_R &= \text{diag} (\hat{M}_{c1}, \hat{M}_{c2}, \hat{m}_{e1}, \hat{m}_{e2}, \hat{m}_{e3}) \, , \\
U_0^\dag \mathcal{M}_N U_0 &= \text{diag} (\hat{M}_{n1}, \hat{M}_{n2}, \hat{M}_{n3}, \hat{M}_{n4}, \hat{m}_{\nu1} , \hat{m}_{\nu2}, \hat{m}_{\nu3}) \, .  
\end{align}
For simplicity we will consider real parameters and limit ourselves to the case where 
only $\mu_1 \neq 0$ ($\mu_2 = \mu_3 = 0$). 
For a more general case see e.g.~\cite{Bisset:1998hy}.
Then the relevant squared mass matrices in the chargino sector read
\begin{multline}
\label{MCTMC}
\mathcal{M}_C^T \mathcal{M}_C = 
\bem
M_2^2 + g^2 v_d^2 /2 & M_2 g v_u/\sqrt{2} + \mu_0 g v_d/\sqrt{2} & 0 & 0 & 0 \cr
M_2 g v_u/\sqrt{2} + \mu_0 g v_d/\sqrt{2} & \mu_0^2 + \mu_1^2 + g^2 v_u^2/2 & m_1 \mu_1 & 0 & 0 \cr
0 & m_1 \mu_1 & m_1^2 & 0 & 0 \cr 
0 & 0 & 0 & m_2^2 & 0 \cr
0 & 0 & 0 & 0 & m_3^2 
\eem \, ,
\end{multline}
which is diagonalized by $U_R$, and 
\begin{multline}
\label{MCMCT}
\mathcal{M}_C \mathcal{M}_C^T = \\
\bem
M_2^2 + g^2 v_u^2/2 & \left(M_2 g v_d + \mu_0 g v_u\right)/\sqrt{2} & \mu_1 g v_u/\sqrt{2}  & 0 & 0 \cr
\left(M_2 g v_d + \mu_0 g v_u\right)/\sqrt{2} & \mu_0^2 + g^2 v_d^2/2 & \mu_0 \mu_1 & 0 & 0 \cr
\mu_1 g v_u/\sqrt{2} & \mu_0 \mu_1 & m_1^2 + \mu_1^2 & 0 & 0 \cr 
0 & 0 & 0 & m_2^2 & 0 \cr
0 & 0 & 0 & 0 & m_3^2 
\eem \, ,
\end{multline}
which is relevant for the determination of $U_L$. 

The $7 \times 7$ neutralino mass matrix is given in \eq{LN},  
with $\mu_2 = \mu_3 = 0$ and $\eta_\alpha = 0$. We neglect the contribution 
of the type I + III seesaw, since it can be made subleading (cf.~\sect{seesawGUT}).  

Working in the phenomenological limit $M_{1,2} \approx \mu_{0,1} \gg v_{u,d} = \mathcal{O} (m_W) \gg m_{1}$, at the 
first order in the expansion parameter $\epsilon = m_W / M_{1,2}$ or $m_1 / m_W \ll 1$ we find: 
\begin{itemize}
\item Chargino sector: 
\begin{align}
\hat{M}_{c1} (\epsilon)  & = M_2 \, , \\
\hat{M}_{c2} (\epsilon)  & = \mu \, , \\
\hat{m}_{1} (\epsilon)  & = m_1 \sqrt{1- \frac{\mu_1^2}{\mu^2
%+\frac{g^2v_u^2}{2}
}} \, , \\
\hat{m}_{2} (\epsilon)  & = m_2 \, , \\
\hat{m}_{3} (\epsilon)  & = m_3 \, , 
\end{align}
where $\mu = \sqrt{\mu_0^2 + \mu_1^2}$. The perturbed eigenvectors (normalized up to $\mathcal{O}(\epsilon^2)$ corrections) read
\begin{align}
\tilde{W}^+ (\epsilon) &= \tilde{W}^+ + \frac{g v_u}{\sqrt{2} M_2} \tilde{H}^+_u \, , \\
\tilde{H}^+_u (\epsilon) &= -\frac{g v_u}{\sqrt{2} M_2} \tilde{W}^+ + \tilde{H}^+_u
+ \frac{\mu_1 m_1}{\mu^2
%+\frac{g^2v_u^2}{2}
} e^c_1 \, , \\
e^c_1 (\epsilon) &= - \frac{\mu_1 m_1}{\mu^2
%+\frac{g^2v_u^2}{2}
} \tilde{H}^+_u + e^c_1 \, , \\
e^c_2 (\epsilon) &= e^c_2 \, , \\
e^c_3 (\epsilon) &= e^c_3 \, ,
\end{align}
and
\begin{align}
\label{mixwhd}
\tilde{W}^- (\epsilon) &= \tilde{W}^- + \frac{g v_d}{\sqrt{2} M_2} \tilde{H}^-_d \, , \\
\tilde{H}^-_d (\epsilon) &= - \frac{g v_d}{\sqrt{2} M_2} \tilde{W}^- + \frac{\mu_0}{\mu} \tilde{H}^-_d + \frac{\mu_1}{\mu} e_1 \, , \\
\label{e1eps}
e_1 (\epsilon) &= \frac{g v_d \mu_1}{\sqrt{2} \mu M_2} \tilde{W}^- -\frac{\mu_1}{\mu} \tilde{H}^-_d + \frac{\mu_0}{\mu} e_1 \, , \\
e_2 (\epsilon) &= e_2 \, , \\
e_3 (\epsilon) &= e_3 \, .
\end{align}
Notice that while the mixing between $e^c_1$ and $\tilde{H}^+_u$ is tiny, 
the states $e_1$ and $\tilde{H}^-_d$ have a large mixing angle, i.e.~$\theta_E^1 = \arctan\mu_1/\mu_0 \approx 67^{\circ}$, for the  
required value of $y_1 = \mu_1/\mu_0$ needed to fit the electron mass (cf.~\eq{y1}). 

\item Neutralino sector: 
for the eigenvalues we obtain
\begin{align}
\hat{M}_{n1} (\epsilon) & = M_1 \, , \\
\hat{M}_{n2} (\epsilon) & = M_2 \, , \\
\hat{M}_{n3} (\epsilon) & = -\mu 
%+ \frac{g'^2(\mu v_u - \mu_0 v_d)^2}{8\mu^2 M_1} + \frac{g^2(\mu v_u - \mu_0 v_d)^2}{8\mu^2 M_2} 
\, , \\
\hat{M}_{n4} (\epsilon) & = \mu 
%+ \frac{g'^2(\mu v_u + \mu_0 v_d)^2}{8\mu^2 M_1} + \frac{g^2(\mu v_u + \mu_0 v_d)^2}{8\mu^2 M_2} 
\, , \\
\hat{m}_{\nu1} (\epsilon) &= - \frac{\mu_1^2 v_d^2}{4 \mu^2} \left( \frac{g'^2}{M_1} + \frac{g^2}{M_2} \right) \, , \\
\hat{m}_{\nu2} (\epsilon) &= 0  \, , \\
\hat{m}_{\nu3} (\epsilon) &= 0 \, . 
\end{align}
while, for the eigenstate associated with the massive neutrino (the remaining eigenstates 
are phenomenologically less important and can be easily 
inferred from the relevant mass matrix) we get 
\beq
\label{nu2eps}
\nu_1 (\epsilon) = - \frac{g' v_d \mu_1}{2 \mu M_1} \tilde{B}^0 + \frac{g v_d \mu_1}{2 \mu M_2} \tilde{W}^0 
- \frac{\mu_1}{\mu} \tilde{H}^0_d + \frac{\mu_0}{\mu} \nu_1 \, .  
\eeq
\end{itemize}
The massive neutrino is hence maximally mixed with the neutral Higgsino. 
This is in full analogy with the electron--charged Higgsino mixing in \eq{e1eps}.  
In fact, the source of R-parity breaking $\mu_1$ 
is associated with an SU(2) invariant operator, so we expect the same large mixing 
for both the components of the SU(2) multiplets $L_1 = (\nu_1, e_1)^T$ and $\tilde{H}_d = (\tilde{H}_d^0, \tilde{H}_d^-)^T$.

\end{document}